\documentclass{aa}
\input epsf
\begin{document}

\thesaurus{07         
          (03.13.7,   
           08.01.1,   
           08.01.3,   
           08.03.1,   
           09.04.1,   
           13.09.4)}  

\title{Infrared spectra of meteoritic SiC grains} 

\subtitle{}

\author{A.\,C.\,Andersen\inst{1} \and C.\,J\"{a}ger\inst{2} \and
H.\,Mutschke\inst{2} \and A.\,Braatz\inst{2,3} \and D.\,Cl\'{e}ment\inst{2} \and 
Th.\,Henning\inst{2} \and U.\,G.\,J{\o}rgensen\inst{1} \and U.\,Ott\inst{3}}

\offprints{Anja C.\,Andersen}

\institute{
 Copenhagen University Astronomical Observatory, 
         Juliane Maries Vej 30, DK--2100 Copenhagen, Denmark\\
         email: anja/uffegj@stella.nbi.dk
\and
 Astrophysical Institute and University Observatory (AIU),
         Schillerg{\"a}{\ss}chen 3, D-07745 Jena, Germany\\
         email: conny/mutschke/braatz/dacapo/henning@astro.uni-jena.de
\and
 Max Planck Institut for Chemistry, Becherweg 27, D-55128 Mainz,
         Germany \\ 
         email: braatz/ott@mpch-mainz.mpg.de}

\date{Received July 15, 1998; accepted December 1, 1998}

\maketitle


\begin{abstract}

We present here the first infrared spectra of meteoritic SiC grains. The
  mid-infrared transmission spectra of meteoritic SiC grains isolated
 from the Murchison
  meteorite were measured in the wavelength range 2.5--16.5
  $\mu$m, in order to make available the optical properties of presolar SiC
  grains. These grains are most likely stellar condensates with an
  origin predominately in carbon stars. 
Measurements were performed on two different extractions of presolar
SiC from the Murchison meteorite. The two samples show very different
spectral appearance due to different grain size distributions.  
The spectral feature of the smaller meteoritic SiC grains is a relatively
broad absorption band found between the longitudinal and transverse lattice vibration
modes around 11.3 $\mu$m, supporting the
current interpretation about the presence of SiC grains in carbon
stars.  In contrast to this,
the spectral feature of the large ($>$ 5 $\mu$m) 
grains has an extinction minimum around 10 $\mu$m. 
  The obtained spectra are compared with commercially available SiC grains 
  and the differences are discussed. 
  This comparison shows that the crystal structure (e.g., $\beta$-SiC 
  versus $\alpha$-SiC)
  of SiC grains plays a minor role on the optical signature of SiC
  grains compared to e.g. grain size. 
  

\end{abstract}

%

\section{Introduction}

Following the thermodynamical equilibrium calculations of Friedemann
(1969ab) and Gilman (1969) which suggested that silicon carbide (SiC)
particles could form in the mass outflow of carbon stars, Hackwell
(1972) and Tref\-fers \& Cohen (1974) performed infrared spectroscopy of
such stars and thereby provided the empirical evidence for the presence of
SiC particles in stellar envelopes.  A broad infrared emission
feature seen in the spectra of many carbon stars, peaking between 11.0
and 11.5 $\mu$m is therefore attributed to solid SiC particles and SiC
is believed to be a significant constituent of the dust around carbon
stars.

Presolar SiC grains have been identified in primitive meteorites
(Bernatowicz et al.\ 1987).  Based on isotopic measurements of the
major and trace elements in the SiC grains and on models of stellar
nucleosynthesis, it is established that a majority of the presolar SiC
grains has their origin in the atmospheres of late-type carbon-rich
stars (Gallino et al.\ 1990, 1994; Hoppe et al.\ 1994). For recent reviews
see, e.g., Anders \& Zinner (1993), Ott (1993) and Hoppe \& Ott
(1997).  The grain sizes of presolar SiC from the Murchison meteorite
have been found by Amari et al.\ (1994) to vary from less than 0.05 to
20 $\mu$m in equivalent spherical diameter, with about 95\% (by mass)
of the grains being between 0.3 and 3 $\mu$m. This distribution is
coarser than for presolar SiC found in other meteorites (Russel et
al.\ 1993, 1997; Huss \& Lewis 1995; Gao et al.\ 1996).
Therefore, Russel et al.\ (1997) have speculated that the finer
grained SiC was lost through size sorting in the solar nebula prior to
accretion of Murchison.  In the carbonaceous chondrites only 
approximately 0.004\%
of the silicon is in the form of SiC (the remainder being in the form
of silicates).

Silicon carbide occurs in a large variety of crystal types.  The basic
units are Si-C bilayers with a three-fold symmetry axis, in which the
Si and C atoms are closely packed.  It is the special stacking of
these layers that determines the occurrence of the polytypes.  The
second bilayer is shifted in the [$\bar{1}100$]-direction by
$1/\sqrt{3}$ of the Si-Si or C-C atomic distance in the layer.  If a
third and a fourth layer is stacked in an identical way, then the
atoms in the fourth layer lie exactly above the ones in the first
layer. Further repetition of this sequence results in a cubic crystal
structure called $\beta$-SiC. 
If at least one bilayer is shifted in the opposite direction the resulting 
structure is hexagonal or rhombohedral (Mutschke et al.\ 1999). All 
the polytypes resulting from non-cubic stacking sequences are summarized 
in the term $\alpha$-SiC. 

Virag et al.\ (1992) have investigated
the crystal structure of the large presolar SiC grains 
extracted by Amari et al.\ (1994) (LS and LU series).
The authors investigated forty-one large (from
$1 \times 1.5$ $\mu$m to $23 \times 23$ $\mu$m) grains from the
Murchison meteorite by Raman spectroscopy.  Thirty-two of these grains
were found to have a cubic crystallographic structure ($\beta$-SiC),
the remaining grains showed a non-cubic structure (hexagonal or
rhombohedric; $\alpha$-SiC). However, the $\alpha$-SiC
grains were also characterized by a
normal isotopic composition, indicating that they might not be of
presolar origin.  
Recently Daulton et al.\ (1998) investigated one of the finer grained samples
extracted by Amari et al.\ (1994) (KJB grain size 0.3$-$0.7 $\mu$m) and
found that for these smaller grains there seems to be even amounts of
$\alpha$- and $\beta$-SiC.  This indicates that while the larger presolar
grains seem to be dominated by the $\beta$-SiC type the smaller grains are
a mix of the $\alpha$- and $\beta$-SiC type, but the presolar nature of
the small $\alpha$-SiC grains still needs to be confirmed.
Whether presolar SiC grains will turn out to be of predominately one or the
other crystal type can place constraints on the formation parameters of the 
grains. Therefore, it has been attemted by several groups to derive the 
crystal type of circumstellar SiC grains from their observed IR emission 
spectra (e.g. Blanco et al.\ 1994, 1998; Groenewegen 1995; Speck et al.\ 1997). However, it is argued by Papoular et al.\ (1998) that the
IR band profiles might not be sensitive to the crystal type but to other
structural and morphological grain properties. To decide these questions, more
laboratory studies are needed (Mutschke et al.\ 1999).

We have previously published the absorption coefficients of presolar diamonds
(Mutschke et al.\ 1995; Andersen et al.\ 1998) and along this line of providing
the spectral feature of ``real star dust'', we have now measured the
spectral appearance of presolar SiC grains.  These data are necessary 
in order to determine if the spectral 
emission (and in a few cases absorption (Jones et al.\ 1978; Speck et
al. 1997)) feature found in circumstellar envelopes of carbon stars,
is consistent with the meteoritic grains having
originated in such stars. Agreement between the isotopic composition of the
presolar SiC grains and those predicted in AGB star nucleosynthetic models,
give strong belief that these meteoritic grains originated in carbon stars.
It is our hope that comparison between the optical properties of presolar
SiC and the appearance of the dust features in carbon star spectra will be
able to impose further constraints.
In this paper, we present results from measurements of
the spectral properties of
meteoritic SiC in the wavelength range between 2.5 and 16.5 $\mu$m.

\section{The extraction procedure}

Two fractions of 26.8 g (sample~{\sc i}) and 10.0 g (sample~{\sc ii})
from an originally large (100 g) piece of Murchison
(a CM2 chondrite) was used for the extraction of meteoritic SiC grains.  The
originally large piece was crushed with a steel mortar, to obtain
smaller fractions.

To obtain a large and clean sample of the SiC grains for the
spectroscopic measurement, it is necessary to isolate the presolar
grains from the rest of the meteoritic material.  A physical,
non-destructive separation does not work well because the grains are
tiny and cling to the much larger amounts of fine-grained clay
minerals and kerogen (macromolecular organic matter). Therefore, a
destructive chemical separation in which undesirable minerals are
dissolved by appropriate reagents is necessary.  The chemical
separation procedures used in this study were variants of those
developed by Tang \& Anders (1988) and Amari et al.\ (1994),
which mainly consist of a set of progressively more corrosive ``selective
solvents'' to remove the mineral phases one by one, but also includes density
separations.  

Extracting presolar SiC from meteorites is much more complicated than
extracting the presolar nano-diamonds, because the SiC (6 ppm) 
is much rarer than the nano-diamonds (500 ppm) (Amari et al.\ 1994), and
because the diamonds can be density separated out of the meteoritic
sample without having to remove all mineral phases first.  
It is therefore much harder to obtain a clean sample
of presolar SiC grains than of presolar diamonds.  
There will in most cases always be smaller
amounts of corundum, hibonite, spinel and chromite grains present, which are
 very hard to get rid of without also losing the SiC grains.

The extraction treatment was primarily designed to result in a
fraction of very clean presolar nano-diamonds (Braatz in prep.).  The individual
extraction steps for the two samples were almost identical,
except that for the first two steps sample~{\sc i} was treated in a
sealed Teflon bomb at 180$^{\circ}$C while sample~{\sc ii} had a more
gentle treatment without the Teflon bomb and at 80$^{\circ}$C. The
extraction steps were the following: 1) solution of concentrated HCl
(32~\%) to dissolve metals and sulfides; 2) alternating treatment with
concentrated HF (48~\%) and concentrated HCl to dissolve silicates; 
3) extraction
of precipitated sulfur with CS$_{2}$; 4) oxidation with 0.5\,N
Na$_{2}$Cr$_{2}$O$_{7}$ + 2\,N H$_{2}$SO$_{4}$ at 80$^{\circ}$C; 5)
extraction of the nano-diamonds as a colloid with a solution of bidistilled
water and NH$_{3}$ (pH 9--10).  This step results in a
precipitate at the bottom of the container
containing mainly residual spinels and SiC.  
The nano-diamonds were used for other
studies (Braatz in prep.).  The spinel-SiC residue was further
treated by; 6) boiling in 70\% HClO$_{4}$ (200$^{\circ}$C) to remove
residual organic material and graphite; 7) boiling in H$_{2}$SO$_{4}$
(180 $^{\circ}$C) to remove the spinels (spinel MgAl$_{2}$O$_{4}$ and
chromite FeCr$_{2}$O$_{4}$) and 8) another treatment with HF/HCl
(60$^{\circ}$C) to remove remaining silicon bearing grains other than
SiC.

Most of the treatments were carried out in Teflon containers.  Between
all steps intensive washing was carried out, with diluted HCl (pH $<$
2) and/or bidistilled water, which is important in order to remove
possible coatings that impede further reactions (Amari et al.\ 1994).
Each time the supernatant liquid was pipetted into separate discard 
tubes for the two samples and all solids which settled down in the 
discard tubes were returned to the main sample.

\section{Spectroscopy}

\subsection{Measurement procedures}

Several groups have carried out experimental work on the spectroscopic
properties of commercially available and laboratory produced SiC particles. 
Stephens (1980) has studied laser-produced 
$\beta$-SiC condensates, Friedemann et al.\ (1981) measured spectra of 
commercially available $\alpha$-SiC, Borghesi et al.\ (1985) studied
commercially produced $\alpha$-SiC and $\beta$-SiC, Kaito et al.\ (1995) 
have studied $\alpha$-SiC and $\beta$-SiC produced by simultaneous 
evaporation of silicon and carbon and Papoular et al.\ (1998) 
investigated two samples of $\beta$-SiC powders, one commercially available
and one produced by laser pyrolysis.  

All the groups have done their measurements by embedding
the sub-micrometer particles in a solid matrix eit\-her of 
KBr or CsI.  In the KBr/CsI pellet technique, small quantities of the 
sample are mixed thoroughly with powdered KBr/CsI. Due to the 
softness of the matrix material and its bulk transparency in the mid-IR, 
the material can be pressed into a clear pellet.

According to scattering calculations, embedding the sample in a matrix will  
influence the wavelength at which the frequency-dependent 
extinction falls as well as the intensity, in a way which depends on
the sample considered and the matrix in which it is included (Bohren
\& Huffman 1983; Papoular et al.\ 1998; Mutschke et al.\ 1999). 

Friedemann et al.\ (1981) and Borghesi et al.\ (1985) tried to 
correct the influence of the matrix by blue-shifting the whole 
feature by an amount of $\delta \lambda$ = -0.4~$\mu$m and 
$\delta \lambda$ = -0.1~$\mu$m, respectively. They also corrected the 
intensity by a factor of 0.7 and 0.9, respectively. 
These procedures have been argued by Papoular et al.\ (1998) to be
incorrect, since independent of the matrix material, absorption should 
not fall outside the longitudinal and transverse optical phonon frequencies. 
Instead Papoular et al.\ (1998) proposed a new method for 
computing the expected spectrum for the particles in vacuum, which 
works if the dielectric function of the grains can be described 
by a single Lorentzian oscillator. This certainly is the case for 
SiC (Mutschke et al. 1999). 

Our transmission measurements were performed by placing the
presolar SiC grains on a polished Si substrate. This means that the grains
were mainly but not fully surrounded by air, resulting in much less matrix
effects than if the sample had been embedded in a solid medium.  Samples tend to
cluster both in the KBr pellets
and on a substrate and may do this with different cluster morphologies.
At the moment theory is not able to determine what the optical effect of 
clustering is for SiC grains.
We have not been able to find any systematic change of 
band profiles as a result of clustering.  We have also not found systematic
changes of band profiles related to whether we used a Si, NaCl or a KBr
substrate (refractive index: KBr and NaCl (n=1.5), Si (n=3.4)). 
The samples were mounted on the
Si substrates by dispersing them in a droplet of chloroform.  All the
spectroscopic measurements were made with an infrared microscope
attached to a Bruker 113v Fourier Transform Infrared  Spectrometer.
The detector is a liquid-nitrogen-cooled mercury cadmium telluride
detector with a spectral range of 7000--600~cm$^{-1}$. The sampling
diameter of the microscope can be as small as 30 $\mu$m.  Spectra were
obtained on different grain clusters of the samples which were 10 to
80~$\mu$m in size. The microscope aperture used for the measurements
was always 80~$\mu$m since this gave a sufficient signal-to-noise
ratio with a reasonable number of scans (64).  The measurements were
performed with a resolution of 1~cm$^{-1}$.  The reference spectra
were obtained on a blank part of the substrate.

\subsection{Results}

\begin{figure}
  \centering \leavevmode \epsfxsize=1.00 \columnwidth
  \epsfbox{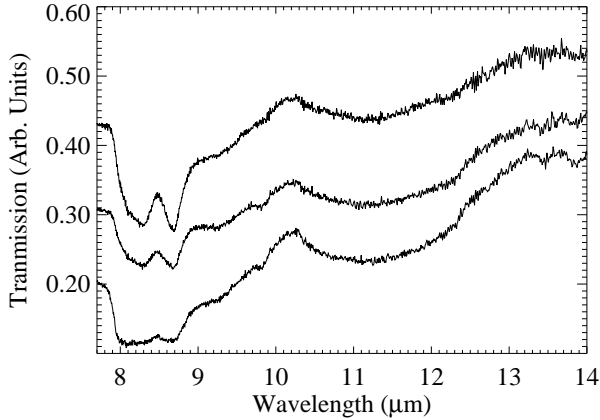}
  \caption[]{Infrared spectra of 
    meteoritic SiC grains (sample~{\sc i}) from the Murchison meteorite
    obtained with the IR microscope at an aperture of 80 $\mu$m. The
    SiC feature is located at about 11.3 $\mu$m. The
    features around 8.2 and 8.6 $\mu$m are due to Teflon (see
    text).}
  \label{mj}
\end{figure}

\begin{figure}
  \centering \leavevmode \epsfxsize=1.00 \columnwidth
  \epsfbox{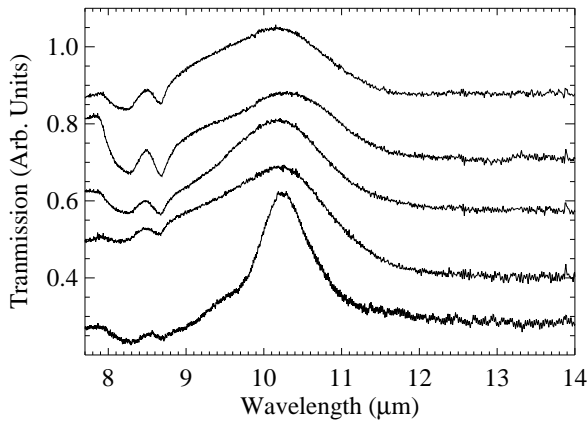}
  \caption[]{Infrared spectra of 
    meteoritic SiC grains (sample~{\sc ii}) from the Murchison meteorite
    obtained with the IR microscope at an aperture of 80 $\mu$m. The
    appearance of the feature is influenced by the Christiansen
    effect (see text).}
  \label{m1718}
\end{figure}

One of the drawbacks with using the infrared microscope is that a
reliable mass estimate of the fraction responsible for the spectral
feature cannot be obtained.  With the microscope one measures
different parts of the sample and depending on how good the sample was
mounted the mass can be more or less evenly distributed on the
substrate.  
Different densities will result in different depths of the
features. The results shown in Fig.\,\ref{mj} for sample~{\sc i} and
in Fig.\,\ref{m1718} for sample~{\sc ii} are the raw data (y is shifted
for better comparison).  There is a
remarkable difference between the spectral appearance of the 
SiC grains in the two samples, despite the fact that they were prepared
by using almost identical extraction procedures and came from the same
larger piece of meteorite.  This difference can be explained by
different grain sizes (see later).

Figure\,\ref{mj} (sample~{\sc i}) shows the expected broad absorption 
feature of small SiC grains between the positions of the longitudinal 
and transverse lattice vibration modes at $\omega_L\approx$ 10.3~$\mu$m
and $\omega_T\approx$ 12.6~$\mu$m (Mutschke et al.\ 1999). 
The center of the band is located at about 11.3 $\mu$m.  
The peaks around 8.2 and 8.6 $\mu$m are due to 
--CF and --CF$_2$ groups originating from the treatment in Teflon containers 
with hot concentrated sulphuric acid (H$_{2}$SO$_{4}$). The sulphuric acid did not only 
dissolve spinel grains but also attacked the walls of the container.

As an attempt to try to get rid of the Teflon-related contaminate, we
considered a density separation of the Teflon and the SiC grains by use of e.g
sodium-polywolframate
(3Na$_{2}$WO$_{4}\cdot$9WO$_{3}\cdot$H$_{2}$O) (Amari et al.\ 1994). However,
all possible solvents, that we could think of, tended to have IR features 
in exactly the same infrared spectral region as the SiC feature we were 
interested in. Therefore, we did not follow this approach.

In Fig.\,\ref{m1718} (sample~{\sc ii}) there is apparently no 
similar SiC band. Instead one observes a maximum of the transmission 
spectrum more or less at the place of the longitudinal lattice 
vibration mode. At longer wavelengths the spectrum is flat. In the 
next section we will explain that this does not mean the absence of SiC grains.

\subsection{Size effects}

Due to collective processes very small solid particles (small 
compared to the wavelength) exhibit strong
resonances in absorption in the spectral regions where the real part 
of the dielectric function ($\epsilon_{1}$) is negative.  
The precise positions of these resonances, called surface modes,
depend on the particle shape, size,
and on the nature and amount of coatings or matrixes surrounding the grain 
(Bohren \& Huffman 1983).  There are two distinct energy ranges in which
resonances occur.  One is in the infrared in the region of strong
lattice bands between the transverse optical phonon frequency
($\omega_{T}$) and the longitudinal optical phonon frequency
($\omega_{L}$).  The other is in the ultraviolet and is due to 
the transitions of bound electrons. 

For the surface modes of very small SiC particles in the infrared, 
Gilra (1973) performed Mie calculations for different shapes 
and found that for a thin disc the resonances are at
10.3 $\mu$m ($\omega_{L}$) and 12.6 $\mu$m ($\omega_{T}$).  As the
oblateness decreases, the resonances move towards each other and
finally for the spherical case there is only one resonance at 10.73
$\mu$m ($\epsilon_{1} = -2$).  As the particles become prolate, the
resonances move away from each other and finally for the case of a
needle they are at 12.6 $\mu$m ($\omega_{T}$) and 10.55 $\mu$m
($\epsilon_{1} = -1$).  Gilra (1973) concludes that if the particles
are highly irregular there will be a broad feature between about 10.3
$\mu$m and 12.6 $\mu$m, which agrees very well with the later findings
of Bohren \& Huffman (1983) resulting in the continuous distribution of
ellipsoids (CDE) approximation. 

If the grain size is larger than or comparable to the wavelength, 
scattering becomes more important and diminishes the transmitted 
light also at frequencies outside the absorption band. 
At a special (larger) frequency outside but close to the absorption band, 
$n \simeq 1$ and $k \simeq 0$. Since these values are close to those of 
the surrounding air (or free space) then at this spectral point the particles 
are nearly invisible. In other words, scattering is greatly reduced 
and a maximum in the transmittens spectrum is observed. This is known as the 
Christiansen effect (Bohren \& Huffman 1983).

\begin{figure}
  \centering \leavevmode \epsfxsize=1.00 \columnwidth 
    \epsfbox{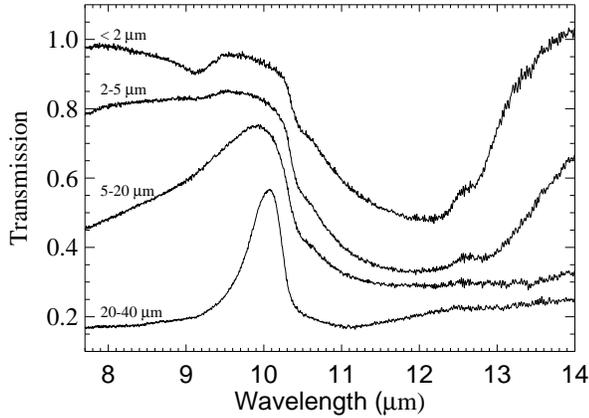}
  \caption[]{Spectral appearance of commercially available $\alpha$-SiC 
    (Duisburg) sedimented to obtain different grain size fractions.
    The variation of the spectral appearance as a result of different
    grain size is apparent. As the grain size increases the influence
    of the Christiansen effect becomes important (see text).}
 \label{size}
\end{figure}

\begin{figure}
  \centering \leavevmode \epsfxsize=1.00 \columnwidth 
  \epsfbox{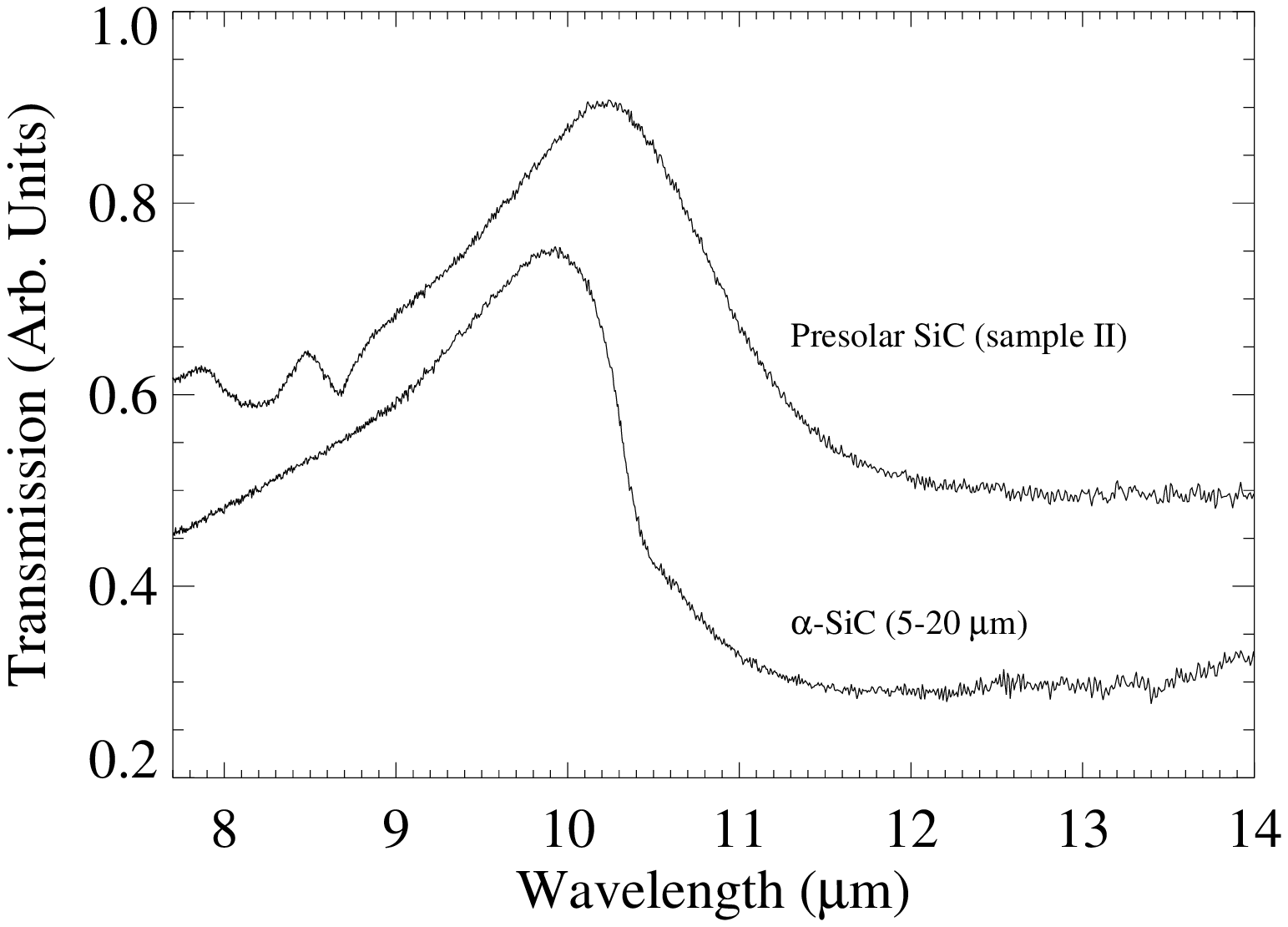 }
  \caption[]{Spectra of meteoritic SiC grains (sample~{\sc ii}) and 
commercially available $\alpha$-SiC (Duisburg) with grain sizes 5--20 $\mu$m.}
 \label{sample2}
\end{figure}

\begin{figure}
  \centering \leavevmode \epsfxsize=1.00 \columnwidth
  \epsfbox{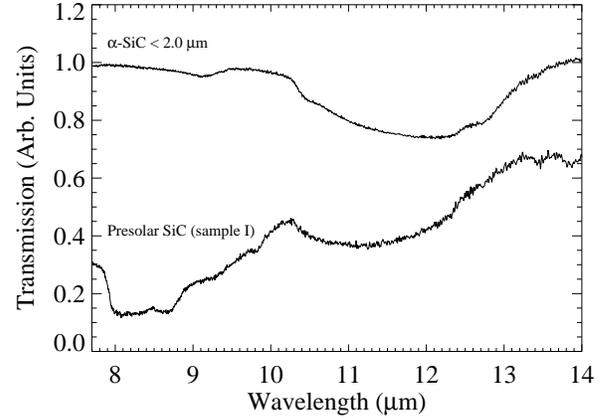}
  \caption[]{Spectra of meteoritic SiC grains (sample~{\sc i}) and commercially available 
    $\alpha$-SiC (Duisburg) with grain sizes less than 2$\mu$m.}
 \label{sample1}
\end{figure}

In order to investigate these size effects we did infrared microscope 
measurements on a
commercially available $\alpha$-SiC grain sample (Duisburg) with large grain
size of up to 40 $\mu$m.  This sample was sedimented in acetone and 
based on settling rates calculated from Stokes law
\begin{equation}
t = 18 \frac{h \cdot \mu_{dyn}}{g \cdot d^{2} \cdot (\rho_{grain} - \rho_{liquid})}
\end{equation}
different size fractions were obtained.  Here $t$ is the settling
time, $\mu_{dyn}$ is the dynamic viscosity, $h$ the setteling height, $g$
the gravitational constant, $d$ the grain diameter, $\rho_{grain}$ the
density of the grains and $\rho_{liquid}$ the density of the liquid.
The size fractions that were obtained were 40--20 $\mu$m, 20--5 $\mu$m, 
5--2 $\mu$m and $<$ 2 $\mu$m.  These samples have been mounted on 
Si substrates in the same way as for the meteoritic samples. 

In Fig.\,\ref{size} it is shown that the
spectral appearance of SiC grains strongly depends on the grain
size and that the Christiansen effect dominates for the larger
grains. This means that the absorption feature between
$\omega_{T}$ and $\omega_{L}$ is transformed into a flat spectrum with 
a transmission maximum at about $\omega_{L}$ (see above).

Figure\,\ref{sample2} shows sample~{\sc ii} of the meteoritic SiC compared to 
the $\alpha$-SiC (Duisburg)  sample with grain sizes  5--20 $\mu$m.  
Figure\,\ref{sample1} shows sample~{\sc i} of the meteoritic
SiC grains compared to the same material but with grain sizes $<$ 2 $\mu$m.  
These comparisons show that the meteoritic SiC grain spectra look
like the profiles expected from SiC grains of different sizes. It may seem
odd that an almost identical extraction procedure on two parts of
originally one piece of the Murchison meteorite should result in two
samples with different size distributions. However, already the results of
Amari et al.\ (1994) indicate an unusual grain size distribution of
extracted Murchison SiC when compared with (less processed) SiC residues
from other meteorites (see discussion in Russel et al.\ 1997). In
addition, we cannot exclude that we lost (presumably mostly the finer)
grains  during our own extensive extraction
procedure. A hint that this may have happened comes from the noble gas
analysis of one of the diamond fractions, in which we found small, but
easily detectable amounts of Ne-E(H), indicating the presence of small
amounts of presolar SiC in the nominal diamond fractions. 

It was not possible to obtain a reliable grain size estimate of the samples
by the use
of a scanning electron microscope (SEM) due to clustering of the grains.
Sample~{\sc i} looks like small grains ($<$ 1 $\mu$m) mixed with Teflon and
some of the more resisting meteoritic grains such as spinel, chromite, hibonite
and corundum, while sample~{\sc ii} looks like larger grains in a cleaner
environment.

The different Christiansen frequencies observed for the meteoritic 
and the commercial SiC grains (Fig.\,\ref{sample2}) hint to different 
optical constants for the two samples. However, so far we have not 
been able to correlate this to a certain polytype of the meteoritic 
SiC grains.
 
The measured spectra of sample~{\sc ii} (Fig.\,\ref{sample2}) 
can therefore be understood as being a spectrum of large meteoritic SiC grains,
where instead of an extinction maximum around 11.3  $\mu$m we see an 
 minimum around 10 
$\mu$m. The measured spectra of sample~{\sc i} (Fig.\,\ref{sample1})
corresponds to a spectrum of mainly small ($<$ 2 $\mu$m) meteoritic grains.

\subsection{Comparison with $\alpha$- and $\beta$-SiC samples}

\begin{figure}
  \centering \leavevmode \epsfxsize=1.00 \columnwidth
  \epsfbox{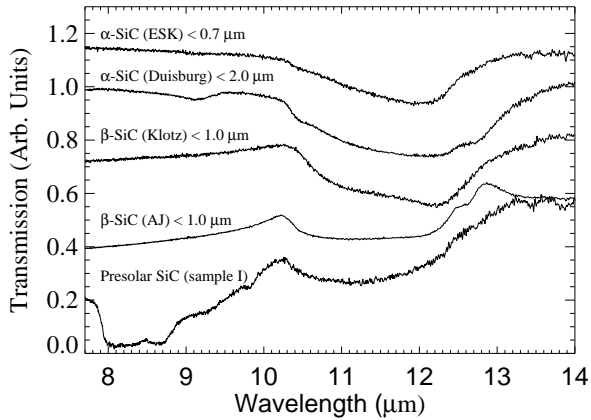}
  \caption[]{Presolar SiC plotted with different commercially available 
    $\alpha$- and $\beta$-SiC samples with the grain sizes indicated. 
    More information about the samples is given in Mutschke et al. (1999).}
 \label{all}
\end{figure}

In the last years, there have been quite a number of papers addressing
the structural differences of SiC particles of different polytype 
as the main important factor influencing the band profile and 
the peak position of the 11.3 $\mu$m feature observed in carbon stars 
(e.g.  Blanco et al.\ 1994, 1998; Groenewegen 1995; Speck et al.\ 1997).  
In another paper (Mutschke et al.\ 1999) we 
study these spectral differences due to the polytype in detail. 
Here, we present infrared microscope measurements on four commercially 
available SiC powders -- two $\alpha$-SiC and two $\beta$-SiC -- 
and compare them to the spectrum of sample~{\sc i} (Fig.\,\ref{all}). 

The five spectra presented in Fig.\,\ref{all} all show the SiC infrared 
band of small grains between $\omega_L\approx$ 10.3~$\mu$m
and $\omega_T\approx$ 12.6~$\mu$m. Obviously, there are significant differences 
in band shape and peak position between the spectra. However, these 
differences are not related to the polytypes. The meteoritic spectrum 
resembles most the $\beta$-SiC (AJ) band profile but this definitely cannot be taken 
as a indication of the polytype of the meteoritic grains as one can 
see by comparing with the spectrum of $\beta$-SiC (Klotz) in Fig.\,\ref{all}. 

This is consistent with the findings of Papoular et al.\ (1998) and
Mutschke et al.\ (1999) that the band profile and consequently the peak 
wavelength 
of the SiC infrared band depends on the distribution of shapes and grain sizes 
rather than crystal type. 
Therefore using the IR spectral feature of different
polytypes to determine whether one of the other crystal type of SiC dominates in
circumstellar outflow, cannot be re\-commended.

The fact that commercially available SiC samples vary so much in
spectral appearance strengthens the importance of studying the
spectral feature of presolar SiC. However, so far it seems
that from the band profiles we will learn rather about grain shape 
and size than about the polytype of extra-solar grains. 
In any case further intensive laboratory studies are needed. 
A step towards a better understanding will be presented in 
Mutschke et al.\ (1999). 

\section{Summary and Conclusions}

The discovery of meteoritic dust grains with an origin outside the
solar system has opened the possibility of studying presolar material
directly in the laboratory. A large fraction of this material is
likely to be dust from the envelopes of asymptotic giant branch (AGB)
stars.  We have previously measured the absorption coefficients of
presolar diamonds (Mutschke et al.\ 1995; Andersen et al.\ 1998).  We
here report the results of mid-IR measurements of meteoritic SiC grains.

Measurements were performed on two different extractions of presolar
SiC from the Murchison meteorite. The two samples show very different
spectral appearances which we interpret as being due to different
grain size distributions in the two extractions.
The spectral feature of the smaller meteoritic SiC grains is at 11.3
$\mu$m, whereas the large ($>$ 5 $\mu$m) SiC grains have no
extinction maximum at 11.3 $\mu$m, but instead are characterized by an
extinction minimum around 10 $\mu$m.
It is a common interpretation that the 11.3$\mu$m band observed in carbon
stars is due to SiC dust. It is also a common interpretation (based on
comparison of isotopic ratios in the meteoritic SiC and nucleosynthesis
models of AGB stars) that the majority of the presolar SiC grains come from 
carbon stars.
If both of these common interpretations are correct,
we conclude from our measurements that the 11.3$\mu$m feature in carbon
stars can be understood as being caused by the smaller end of the size
distribution of SiC of the type identified in primitive meteorites,
and the larger grains must correspond to a grain distribution not yet
identified in carbon stars. 

In the observational data by Speck et al.\ (1997) the (interpreted) SiC
feature in carbon star spectra peak around 11.3 $\mu$m for about 40\%
of cases (13 out of the sample of 30 stars) with nearly symmetric
profiles and a FWHM around 1.8 $\mu$m.  These features, which are relatively
broad just as the feature of the small meteoritic SiC (sample~{\sc i}),
may be interpreted as an indication that the circumstellar SiC is of this
small grain size. If the grain size distribution evolves towards
larger grains, for example during later stages of the carbon star
evolution, it would result in a weakening (and possible disappearance)
of the 11.3 $\mu$m feature, which could explain the remaining 60\% of 
stellar spectra. However, there are several strong molecular features 
in the 10--14
$\mu$m area in carbon stars (Hron et al.\ 1998), and a unique
interpretation of the observational data still awaits a
self-consistent simulation taking both the molecular absorption and the
dust emission into account for a wide range of types of carbon stars. 

The fact that large ($>$ 5 $\mu$m) SiC grains have a different spectral
appearance than smaller ($<$ 2 $\mu$m) SiC grains, will make large SiC
grains difficult to observe in the interstellar medium because of the
presence of silicate-related absorption around 10 $\mu$m.  
If the majority of the cosmic SiC grains have the size distribution 
found by Amari et al.\ (1994),
this will probably not significantly change the
abundance limit of less than a few percent SiC
compared to silicates in the ISM (Whittet et al.\ 1990),

The results of spectral measurements on commercially available SiC grain
samples of different polytypes and the meteoritic SiC grain samples
show that the variations among polytypes of SiC
grains are smaller than the variations due to different grain size.
It is therefore not possible to distinguish, by IR spectroscopy, between
$\alpha$- and $\beta$-SiC of dusty material as also discussed by
Papoular et al.\ (1998) and Mutschke et al.\ (1999).  

\begin{acknowledgements}            
  
  The authors would like to thank Gabriele Born and Walter Teuschel for
  help with the experiments and valuable discussions.
This work has been supported by the Danish Natural Science Research Council and
by the Deutsche Forschungsgemeinschaft (DFG grant Mu 1164/3-1). 
H.\,M. is supported by a grant from the Max Planck Society to Th.\,H.

\end{acknowledgements}

\end{document}